\documentclass[11pt]{article}

\usepackage{amssymb}
\usepackage{amsmath}
\usepackage{amscd}
\usepackage{bbm}
\usepackage{latexsym}
\usepackage{ifthen}
\usepackage[xdvi]{graphics}
\catcode`@=11
\renewcommand{\section}{\@startsection{section}{1}{0pt}{\medskipamount}
{\medskipamount}{\large\bf}}
\numberwithin{equation}{section}
\catcode`@=12

\def\a{\alpha}
\def\b{\beta}

\def\de{\delta}

\def\k{\kappa}

\def\ps{\psi}

\def\S{\Sigma}

\newcommand{\C}{\mathbbm{C}}
\newcommand{\R}{\mathbbm{R}}
\newcommand{\Z}{\mathbbm{Z}}
\newcommand{\Na}{\mathbbm{N}}

\def\>{\rangle}
\def\<{\langle}
\def\N2{N=2}
\def\pa{\partial}

\def\sfrac#1#2{{\textstyle\frac{#1}{#2}}}

\newcommand{\ab}{{\bar{a}}}

\newcommand{\kb}{\bar{k}}

\newcommand{\ub}{\bar{u}}

\newcommand{\Zb}{\bar{Z}}

\newcommand{\ov}[1]{\overline{#1}}

\newcommand{\Gt}{\widetilde{G}}
\newcommand{\Ut}{\widetilde{U}}
\newcommand{\Vt}{\widetilde{V}}
\newcommand{\kt}{\tilde{k}}

\newcommand{\ic}{\text{i}}
\DeclareMathOperator{\sn}{sn}
\DeclareMathOperator{\cn}{cn}
\DeclareMathOperator{\dn}{dn}

\newcommand{\Gp}{\ifthenelse{\boolean{mmode}}{{G^+}}{\mbox{$G^+\:$}}}
\newcommand{\Gtp}{\ifthenelse{\boolean{mmode}}{\mbox{$\Gt^+$}}{\mbox{$\Gt^+\:$}}}
\newcommand{\Gm}{\ifthenelse{\boolean{mmode}}{{G^-}}{\mbox{$G^-\:$}}}
\newcommand{\Gtm}{\ifthenelse{\boolean{mmode}}{\mbox{$\Gt^-$}}{\mbox{$\Gt^-\:$}}}

\newcommand{\uv}{\text{$|\;\!\!\!\uparrow\>$}}
\newcommand{\dv}{\text{$|\;\!\!\!\downarrow\>$}}

\newcommand{\dvb}{\text{$\<\downarrow\:\!\!\!|$}}

\newcommand{\uudvb}{\<\uparrow\uparrow\downarrow\;\!\!\!|}
\newcommand{\uduvb}{\<\uparrow\downarrow\uparrow\;\!\!\!|}
\newcommand{\duuvb}{\<\downarrow\uparrow\uparrow\;\!\!\!|}

\newcommand{\at}{\tilde{a}}
\begin{document}
\begin{titlepage}
\setcounter{page}{0}
\begin{flushright}
{\tt hep-th/0408245}\\
MIT--CTP--3535\\
\end{flushright}

\vskip 2.0cm

\begin{center}

{\Large\bf A Note on $\k$-diagonal Surface States}

\vspace{14mm}

{\large Sebastian Uhlmann$\,^*$}
\\[5mm]
{\em Center for Theoretical Physics \\
Massachusetts Institute of Technology \\
Cambridge, MA 02139, USA}\\[7mm]
August, 2004
\end{center}

\vspace{2cm}

%\noindent
%{\sc Abstract:} \\[.7cm]
\begin{abstract}
We classify all twist-even squeezed states in string field theory
which are diagonal in the $\kappa$-basis and simultaneously surface
states. For this purpose, we derive a consistency condition for the
maps defining $\k$-diagonal surface states. It restricts these maps
to a two-parameter family of Jacobi sine functions. Not all of them
are admissible maps for surface states; standard requirements single
out two one-parameter families containing the generalized butterfly
states and the wedge states.
\end{abstract}
\vfill

\textwidth 6.5truein
\hrule width 5.cm
%\vskip.1in
{\small ${}^*$ {\tt uhlmann@lns.mit.edu}}

\end{titlepage}

%%%%%%%%%%%%%%%%%%%%%%%%%%%%%%%%%%%%%%%%%%%%%%%%%%%%%%%%

\section{Introduction}
\noindent
Over the last years, the study of nonperturbative phenomena in
string theory has enjoyed a lively interest. Research in this
direction revealed that string field theory is the language
in which, e.\,g., the decay of unstable D-brane systems can be
described most naturally (for a review of the subject, see~%
\cite{Taylor:2003gn}). Here, the endpoint of the decay process
corresponds to a solution of the equation of motion of the string
field theory on the unstable brane. Up to now, however, no satisfactory
solutions to this equation have been found even in the simplest setting,
i.\,e., Witten's formulation of bosonic string field theory~%
\cite{Witten:1985cc} -- mainly, due to the following two difficulties:

In this theory, the BRST operator~$Q$ plays the role of the kinetic
operator; it involves matter as well as ghost fields which renders it
impossible to study the equation of motion in each sector separately.
As a loophole, Rastelli, Sen, and Zwiebach proposed to consider string
field theory around the tachyon vacuum and to describe D-branes from
this point of view~\cite{Rastelli:2000hv, Rastelli:2001jb}. It could
be shown that, after a singular reparametrization of the worldsheet,
the kinetic operator consists solely out of ghosts~%
\cite{Rastelli:2000hv, Hata:2001sq, Gaiotto:2001ji, Okuyama:2002yr}.
Thus, the equation of motion factorizes into a matter and a ghost part,
the matter part being simply the condition that the matter string field
is a projector of the string field algebra. Amongst other things, this
discovery triggered an investigation of structural aspects of the string
field algebra.

The second difficulty arises from the star product which is used to
multiply string fields~\cite{Witten:1985cc}. Calculations in the customary
discrete oscillator basis~\cite{Gross:1986ia, Gross:1986fk} are notoriously
difficult; however, they may be facilitated by the fact that it is possible
to choose a basis in the string field theory Fock space in which the
Neumann matrices of the interaction vertex are diagonal~\cite{Douglas:2002jm, Belov:2002fp, Arefeva:2002jj, Erler:2002nr, Belov:2002pd, Belov:2003df,
Belov:2003qt, Maccaferri:2003rz, Kling:2003sb, Ihl:2003fw}. In particular
the analysis of algebraic properties of the string field algebra is
simplified in this continuous\footnote{The Witten vertex was first
reformulated in terms of Moyal products in~\cite{Bars:2001ag, Bars:2002nu}
which are constructed from a discrete basis. This basis is equivalent to
the continuous Moyal basis and would therefore also be a convenient tool
for this kind of structural analysis; however, in the present paper, we
choose the continuous basis for consistency with earlier work~%
\cite{Ihl:2003fw}.} Moyal basis. E.\,g., it is possible to classify all
twist-even projectors in certain conformal field theories which are
diagonal in this basis~\cite{Ihl:2003fw} (see also~\cite{Fuchs:2002zz}).
This classification includes many states without immediate geometrical
interpretation. Furthermore, due to its similarity with the Moyal-Weyl
product, the algebraic approach to the star algebra could be useful for
transferring solution-generating techniques from field theory to string
field theory (see, e.\,g., \cite{Lechtenfeld:2002cu, Kling:2002ht,
Kling:2002vi}).

Alternatively, the star product can be computed via conformal field theory
methods~\cite{LeClair:1988sp, LeClair:1988sj}. They lead to a more
geometric understanding of states and their multiplication in string
field theory; in particular, surface states play an important role
in the structural analysis of the star algebra. These are states with
field configurations arising from path integrations over fixed Riemann
surfaces whose boundary consists of a parametrized open string and a piece
with open string boundary conditions. A very fundamental result in this
realm~\cite{Gaiotto:2002kf} was that all such surfaces with the property
that their boundaries touch the midpoint of the open string lead to projector
functionals in the star algebra. Although this class of projectors is quite
large, it may be rewarding to look for projectors without this singular
property which could eventually lead to D-brane solutions with finite
energy densities.

Both methods obviously have their own advantages, and it usually depends
on the respective case which one is given preference. However, it seems that
it should be possible to make rather strong statements concerning states
which can be analyzed particularly well with both methods, i.\,e., those
which are diagonal in the Moyal basis and simultaneously (twist-even)
surface states. The analysis of these states is the aim of the present
note. It turns out that a complete classification of such states is
possible; they comprise the butterfly family and a generalization of the
wedge states (with arbitrary angles). 
Upholding the belief that possible solutions to the equation of motion
of string field theory should have a geometrical interpretation as
surface states,
the additional demand that the states be simultaneously diagonal in the
$\k$-basis (and in this sense particularly easy to handle algebraically)
apparently is very strict. In principle, one could have
imagined that this class included hitherto unknown states which could
have served as candidate solutions to the equation of motion. Our result
seems to imply that we have to look elsewhere for solutions describing
the tachyon vacuum. Apart from that, we find that surface state
projectors which do not satisfy the sufficient condition that their
boundary touches the midpoint of the open string, are particularly rare --
at least in the subsector under consideration, we only encounter the
identity state.

This result is obtained in three steps: First, we derive a consistency
condition on $\k$-diagonal surface states along the lines of earlier
work~\cite{Ihl:2003fw}. This is done in terms of a fermionic first order
system with weights~0 and~1 (but is in principle possible in any
other conformal field theory) and is the subject of section~\ref{sec:cond},
where we also present the Moyal basis for this system. From the technical
point of view, the consistency condition is a differential equation with
deviating arguments on the map $f(z)$ which defines a possible surface
state. Second, we solve this differential equation; the solution turns out
to be a Jacobi sine function parametrized by two complex parameters.
This is the task of section~\ref{sec:solvcc}. Third, we examine
further constraints on these two parameters in section~\ref{sec:kdiagss}
since not all solutions to the consistency condition are eligible as
surface states. This leads to the above-described result. The necessary
background on Jacobian elliptic functions for this analysis is presented
in appendix~\ref{sec:Jac}. Finally, we offer some concluding remarks in section~\ref{sec:concl}.

\section{Derivation of the consistency condition}\label{sec:cond}
\noindent
In this section, we briefly review the derivation of a consistency
condition on twist-even surface states which are diagonal in the
$\k$-basis. Most of the material presented here is a summary of
the relevant facts in~\cite{Ihl:2003fw} which are included in this
note in order to make the discussion self-contained.

The derivation starts from an oscillator representation of the
surface state and leads to a requirement on the maps $f(z)$
defining the surface state; for consistency with the treatment
in~\cite{Ihl:2003fw}, the argument will be presented for a
fermionic first order system with weights $(0,1)$, i.\,e., a
twisted $bc$ system. Naturally, the same condition on $f(z)$ can
be derived for any given conformal field theory.

\noindent
{\bf Continuous basis in the fermionic first order system.} The
fermionic first order system consists of Grassmann-odd fields
$\ps^+$ with weight~0 and $\ps^-$ with weight~1.\footnote{Note
that, w.\,r.\,t.~\cite{Ihl:2003fw}, they are rescaled in order to
have canonically normalized commutation relations, i.\,e., $\ps^+_\text{here}=\frac{1}{\sqrt{2}}\ps^+_\text{there}$ and
$\ps^-_\text{here}=\frac{1}{\sqrt{2}}\ps^-_\text{there}$,
respectively. These fields occur as twisted worldsheet fermions in
string field theory for N=2 strings. Their conformal field theory
coincides with the one common from the construction of solutions to
the ghost part of the equations of motion in vacuum string field theory~%
\cite{Gaiotto:2001ji}; in this context, the fields are customarily denoted
by $c'$ and $b'$. -- Apart from that, the Neumann coefficients $N^{rs}_{kl}$
are rescaled by a factor of~2 w.\,r.\,t.~\cite{Ihl:2003fw}.} As fields of
integral weight, both $\ps^+$ and $\ps^-$ are integer-moded; the modes
satisfy $\{\ps^+_m, \ps^-_n\}=\de_{m,-n}$. In particular, the spin-0
field $\ps^+$ has a zero-mode on the sphere. In analogy to the (untwisted)
$bc$ system there are thus two vacua at the same energy level: the bosonic $SL(2,\R)$-invariant vacuum $|0\>=:\dv$ is annihilated by the Virasoro
modes $L_{m\geq -1}$ and $\ps^+_{m>0}$, $\ps^-_{m\geq 0}$; its fermionic
partner, $\uv:=\ps^+_0\dv$, is annihilated by $\ps^+_{m\geq 0}$,
$\ps^-_{m>0}$. To get nonvanishing fermionic correlation functions, we
need one $\ps^+$-insertion, i.e., $\<\downarrow\!\dv = \<\uparrow\!\uv = 0$, $\<\downarrow\!\uv = 1$.

In terms of these oscillators and vacua, the interaction vertex for the
$\ps^\pm$ system takes the form~\cite{Kling:2003sb} (see also~%
\cite{Maccaferri:2003rz})
\begin{equation}
  \<V_3| = \big( \uudvb+\uduvb+\duuvb \big) \exp\big[ \sum_{r,s}
    \sum_{k=1,l=0}^\infty \ps^{+(r)}_k N_{kl}^{rs} \ps^{-(s)}_l\big]
    \label{eq:psivrtx}
\end{equation}
where the explicit form of the Neumann coefficients $N^{rs}_{kl}$
($1\leq r,s\leq 3$) can be found in~\cite{Kling:2003sb, Ihl:2003fw}. The
(common) eigenvectors of these Neumann matrices are labeled by a
parameter $\k\in\R$; the coefficients of the right and left eigenvectors
$v^+(\k)$ and $v^-(\k)$ are given by the generating functions
\begin{align}
  f_{v_e^-}(\k,z) & = \sum_{n=1}^\infty v_{2n}^-(\k) z^{2n} = \frac{1}
    {\sqrt{{\cal N}(\k)}\k}(1-\cosh\k Z)\, , \label{eq:eogenf1} \\
  f_{v_o^-}(\k,z) & = \sum_{n=1}^\infty v_{2n-1}^-(\k) z^{2n-1} =
    \frac{1}{\sqrt{{\cal N}(\k)}\k}\sinh\k Z\, , \\
  f_{v_e^+}(\k,z) & = \sum_{n=1}^\infty v_{2n}^+(\k) z^{2n} = -\frac{1}
    {\sqrt{{\cal N}(\k)}}\frac{z}{1+z^2}\sinh\k Z\, , \\
  f_{v_o^+}(\k,z) & = \sum_{n=1}^\infty v_{2n-1}^+(\k) z^{2n-1} =
    \frac{1}{\sqrt{{\cal N}(\k)}}\frac{z}{1+z^2}\cosh\k Z \label{eq:eogenf4}
\end{align}
with $Z:=\tan^{-1} z$ and ${\cal N}(\k)=\frac{2}{\k}\sinh\frac{\pi\k}{2}$.
One can show that
\begin{equation}\label{eq:fEVrel}
  \int_{-\infty}^{\infty}d\k\,v_m^+(\k)v_n^-(\k) = \de_{mn}\, , \quad
    \sum_m\,v_m^+(\k)v_m^-(\k') = \de(\k-\k')\, ,
\end{equation}
which can be split into even and odd parts to give
\begin{align}
  2\int_0^\infty d\k\,v_{2m}^+(\k)v_{2n}^-(\k) & = \de_{mn}\, , &
  2\int_0^\infty d\k\,v_{2m+1}^+(\k)v_{2n+1}^-(\k) & = \de_{mn}\, , 
    \label{eq:orth}\\
  2\sum_{n=1}^\infty v_{2n}^+(\k) v_{2n}^-(\k') & = \de(\k-\k')\, , &
  2\sum_{n=1}^\infty v_{2n-1}^+(\k) v_{2n-1}^-(\k') & = \de(\k-\k')
    \label{eq:cmpl}
\end{align}
for $\k>0$. Thus, the continuous modes
\begin{align}
  \psi_{e,\k}^{-\dag} & := \sqrt{2}\sum_{n=1}^{\infty}v_{2n}^-(\k)
    \psi_{-2n}^-\, , &
  \psi_{o,\k}^{-\dag} & := -\sqrt{2}\ic\sum_{n=1}^{\infty}
    v_{2n-1}^-(\k)\psi_{-2n+1}^-\, , \label{eq:contbas1} \\
  \psi_{e,\k}^{+\dag} & := \sqrt{2}\sum_{n=1}^{\infty}v_{2n}^+(\k)
    \psi_{-2n}^+\, , &
  \psi_{o,\k}^{+\dag} & := -\sqrt{2}\ic\sum_{n=1}^{\infty}
    v_{2n-1}^+(\k)\psi_{-2n+1}^+ \label{eq:contbas2}
\end{align}
satisfy the anticommutation relations
\begin{equation}
  \{\psi_{e,\k}^{+\dag},\psi_{e,\k'}^-\}=\,\de(\k-\k')\, ,\qquad
    \{\psi_{o,\k}^{+\dag},\psi_{o,\k'}^-\}=\,\de(\k-\k')
\end{equation}
following from the anticommutation relations of $\psi_n^+$ and
$\psi_n^-$ and the completeness relations of the eigenvectors
$v_n^+(\k)$ and $v_n^-(\k)$. The relations~(\ref{eq:contbas1}) and~%
(\ref{eq:contbas2}) can be inverted:
\begin{align}
  \psi_{-2n}^- & = \sqrt{2}\int_0^{\infty}d\k\; v_{2n}^+(\k)
    \psi_{e,\k}^{-\dag}\, , &
  \psi_{-2n+1}^- & = \sqrt{2}\ic \int_0^{\infty}d\k \; v_{2n-1}^+(\k)
    \psi_{o,\k}^{-\dag}\, , \\
  \psi_{-2n}^+ & = \sqrt{2} \int_0^{\infty}d\k \; v_{2n}^-(\k)
    \psi_{e,\k}^{+\dag}\, , &
  \psi_{-2n+1}^+ & = \sqrt{2}\ic \int_0^{\infty}d\k \;
    v_{2n-1}^-(\k)\psi_{o,\k}^{+\dag}\, .
\end{align}
The interaction vertex~(\ref{eq:psivrtx}) can be rewritten in
the continuous basis; one can take advantage of this fact, e.\,g.,
in order to classify all (twist-even) projectors diagonal in the
$\k$-basis~\cite{Ihl:2003fw}.

\noindent
{\bf Diagonal squeezed states.} In this paper, we restrict the
discussion to squeezed states with definite charge w.\,r.\,t.\
the $U(1)$ current $\ps^+\ps^-$ of the first order system because
only these lead to surface states (recall that this is so since
the Virasoro generators are neutral). In the discrete basis,
such a state takes the form
\begin{equation}
  \<S| = \dvb\exp\Big[ \sum_{m,n}\ps^+_m S_{mn}\ps^-_n \Big] \, ;
  \label{eq:sqstdo}
\end{equation}
in the continuous basis, it features two $\k$-integrations in the
exponent. The $\k$-diagonal states are given by a squeezed state
matrix $S_{mn}$ which commutes with the Neumann matrices~$N^{rs}_{kl}$;
in these cases, the exponent can be reduced to contain only one
$\k$-integration. Namely, for $S_{mn}$, the twist properties of
$v^+_n(\k)$,
\begin{equation}
  v^+_{2n}(-\k) = -v^+_{2n}(\k)\, ,\qquad v^+_{2n+1}(-\k) = v^+_{2n+1}(\k)
    \, ,
\end{equation}
imply that the even and odd parts are also eigenvectors of $S_{mn}$,
\begin{align}
  \sum_n S^{\phantom{+}}_{2m,2n}v^+_{2n}(\k) & = S_{ee}(\k) v^+_{2m}(\k)
  \, , & \sum_n S^{\phantom{+}}_{2m+1,2n}v^+_{2n}(\k) & = S_{oe}(\k)
  v^+_{2m+1}(\k)\, , \notag \\
  \sum_n S^{\phantom{+}}_{2m,2n+1}v^+_{2n+1}(\k) & = S_{eo}(\k) v^+_{2m}
  (\k)\, , & \sum_n S^{\phantom{+}}_{2m+1,2n+1}v^+_{2n+1}(\k) & = S_{oo}
  (\k) v^+_{2m+1}(\k)\, .
\end{align}
Here, $S_{ee}$, $S_{oe}$, $S_{eo}$, and $S_{oo}$, denote the corresponding
eigenvalues.\footnote{Note that for consistency, $S_{ee}$ and $S_{oo}$
have to be even functions of~$\k$, and $S_{eo}$ and $S_{oe}$ have to be
odd functions of~$\k$.} Thus, eq.~(\ref{eq:cmpl}) guarantees that
\begin{equation}
  \sum_{m,n}v^-_{2m}(\k')S^{\phantom{+}}_{2m,2n}v^+_{2n}(\k) = \sfrac{1}
    {2} S_{ee}(\k)\de(\k-\k') \label{eq:vSv}
\end{equation}
and similar relations hold for the other components. If one rewrites
$\ps^+_m S_{mn}\ps^-_n$ in terms of the continuously moded operators,
the delta distributions on the right-hand side of eq.~(\ref{eq:vSv})
can be used to remove the $\k'$-integration. Therefore,
\begin{equation}
  \< S| = \dvb\exp\Big[\frac{1}{2}\int_0^\infty d\k\,\vec{\ps}^+_\k
    \cdot S(\k)\cdot\vec{\ps}^-_\k \Big] \label{eq:sqstate}
\end{equation}
with $S(\k)=\left(\begin{smallmatrix} S_{ee} & -\ic S_{eo}\\ -\ic S_{oe} &
-S_{oo}\end{smallmatrix}\right)$.

\noindent
{\bf Surface states.} A surface state $\<\S^f|$ is determined by a map
$f\colon H\to\S$ from the canonical upper unit half disk $H:=\{|z|\leq 1\}
\subset\C$ onto a Riemann surface~$\S$ with boundary and the requirement
that
\begin{equation}
  \<\S^f|\phi\> = \< f\circ\phi(0)\>_\S \label{eq:surfstdef}
\end{equation}
for all Fock space states~$|\phi\>$ of weight~$h$ in the boundary CFT
with corresponding operators $\phi(z)$. The correlation function
$\<\,\>_\S$ is evaluated on~$\S$, and $f\circ\phi(0)=\big(f'(0)\big)^h
\phi(f(0))$ is the conformal transform of~$\phi$ by the map~$f$.

Now, let us evaluate the correlation function~(\ref{eq:surfstdef}) for
a surface state of the form~(\ref{eq:sqstdo}) with~$|\phi\>=\ps^+(z)
\ps^-(w)\ps^+(0)\dv$,\footnote{The $U(1)$ charges of the insertions of
the correlation functions have to sum up to $+1$ in
order to give a nonvanishing result.}
\begin{equation}
  \dvb\exp\Big[\sum_{m,n>0}\ps^+_m S^{\phantom{+}}_{mn}\ps^-_n\Big]
    \sum_{k=-\infty}^\infty \frac{\ps^+_k}{z^k}\sum_{l=-\infty}^{-1}
    \frac{\ps^-_l}{w^{l+1}}\dv = \sum_{m,n>0} S_{mn}
    z^n w^{m-1} + \frac{1}{z-w}\frac{z}{w}\, .
\end{equation}
On the other hand, according to eq.~(\ref{eq:surfstdef}), this correlation
function should be equal to\footnote{Recall that the second
factor originates from the nontrivial background charge of the system~%
\cite{Kling:2003sb}.}
\begin{equation}
  \< f\circ\phi\>_\S = \frac{f'(w)}{f(z)-f(w)}\frac{f(z)-f(0)}
    {f(w)-f(0)}\, ;
\end{equation}
and we can use $SL(2,\R)$ invariance to fix $f(0)=0$, $f'(0)=1$, and
$f''(0)=0$. Then, the nonsingular part of the correlator becomes~%
\cite{Okuda:2002fj, Bars:2003gu} 
\begin{equation}
  S(z,w) := \sum_{m,n>0} S_{mn} z^n w^{m-1} = \frac{f'(w)}{f(z)-f(w)}
    \frac{f(z)}{f(w)} - \frac{1}{z-w}\frac{z}{w}\, , \label{eq:Szw}
\end{equation}
therefore, deriving w.\,r.\,t.\, $z$ and choosing $w=0$,
\begin{equation}
  \frac{\pa}{\pa z}S(z,0) = -\frac{f'(z)}{f(z)^2} + \frac{1}{z^2}\, .
\end{equation}
This equation can be integrated to give
\begin{equation}
  f(z) = \frac{z}{z S(z,0) + 1} \label{eq:fS}
\end{equation}
for a candidate function~$f$ for the map defining the surface state~%
$\< \S^f|$. Here, the integration constant was chosen in such a way that
for $\< \S^f|=\dvb$, we obtain the identity map $f(z)=z$.

Obviously, the defining map~$f(z)$ is encoded in the coefficients~$S_{mn}$
via~(\ref{eq:Szw}); we can extract a candidate function with the help of
eq.~(\ref{eq:fS}). For a diagonal surface state, the same information can
be equivalently obtained from the eigenvalues $S_{ee}$, $S_{oe}$, $S_{eo}$,
and $S_{oo}$. For our purposes we restrict to twist-even states, i.\,e.\
states with $S_{eo}=S_{oe}=0$. Now, $S(z,w)$ can be reconstructed
from these eigenvalues using~(\ref{eq:orth}), (\ref{eq:eogenf1})--%
(\ref{eq:eogenf4}), and~(\ref{eq:vSv}),
\begin{equation}
  S(z,w) = \frac{2}{w}\int_0^\infty d\k\Big[ f_{v_e^+}(\k,w)f_{v_e^-}
   (\k,z) S_{ee}(\k) + f_{v_o^+}(\k,w)f_{v_o^-}(\k,z)S_{oo}(\k)\Big]\, .
  \label{eq:Szw2}
\end{equation}
It should be noted that $f(z)$ is purely given by the $S_{oo}$-part, see~%
(\ref{eq:fS}) and
\begin{equation}
  S(z,0) = \int_0^\infty d\k\,\frac{\sinh \k Z}{\sinh\frac{\pi\k}
    {2}}S_{oo}(\k)\, . \label{eq:Sz0int}
\end{equation}
As above, $Z$ abbreviates~$\tan^{-1} z$. Obviously, $f$ is an {\em
odd} function mapping the upper unit half-disk into some region of the
upper half-plane.

\noindent
{\bf Consistency condition for surface states.} We will now derive a
necessary condition on the maps $f(z)$ defining a surface state. This
condition consists in a differential equation with deviating arguments,
which, however, turns out to be exactly solvable. In this way, we
will show that all twist-even surface states belong to a two-parameter
class of states. Not all functions in this two-parameter family define
surface states; further restrictions single out the butterfly family
and the wedge states with arbitrary angles (including the identity).

The starting point for the derivation of the condition is the observation
that the $S_{oo}$-part of eq.~(\ref{eq:Szw2}) is given by $S(z,0)$ and
the $S_{ee}$-part can be determined via (an integral of) $\frac{\pa}{\pa w}S(z,w)|_{w=0}$. Thus, the full integral~(\ref{eq:Szw2}) can be computed
via~$f(z)$. Together with~(\ref{eq:Szw}), this gives a restriction
on~$f(z)$.

To begin with, note that~(\ref{eq:Szw2}) can be rewritten as
\begin{equation}
  S(z,w) = \frac{1}{1+w^2}\Big[-S_1(W) + \frac{1}{2}S_1(Z+W) -
    \frac{1}{2}S_1(Z-W) + \frac{1}{2}S_2(Z+W) + \frac{1}{2}S_2(Z-W)
    \Big]\, , \label{eq:Szwsplit}
\end{equation}
again with $W=\tan^{-1} w$, $Z=\tan^{-1} z$, and
\begin{equation}
  S_1(Z) = \int_0^\infty d\k\,\frac{\sinh \k Z}{\sinh\sfrac{\pi\k}{2}}\,
    S_{ee}(\k)\, ,\qquad
  S_2(Z) = \int_0^\infty d\k\,\frac{\sinh \k Z}{\sinh\sfrac{\pi\k}{2}}\,
    S_{oo}(\k)\, .
\end{equation}
Obviously, $S_2(\tan^{-1} z)=\frac{1}{f(z)}-\frac{1}{z}$.

Next, it is easy to see that due to the form of~(\ref{eq:Szw2}), only
the $S_{ee}$-part will contribute to $\tilde{S}(Z):=\frac{\pa}{\pa w}
(1+w^2)S(z,w)|_{w=0} = \frac{\pa}{\pa w}S(z,w)|_{w=0}$; in addition, one can
recover $S_1(Z)$ fully from $\int_0^Z dZ' (\tilde{S}(Z') + \tilde{S}_0)$
with some additive constant~$\tilde{S}_0$. Eq.~(\ref{eq:Szwsplit}) ensures
that the contributions of this constant to $S(z,w)$ cancel.

Following this prescription, we have to compute $\frac{\pa}{\pa w}
S(z,w)|_{w=0}$ from~(\ref{eq:Szw}). A direct computation with our choice
$f(0)=0$, $f'(0)=1$, $f''(0)=0$ leads to a difference of singularities.
In order to avoid this one can compute $\frac{\pa^2}{\pa z\pa w}S(z,w)$,
set $w=0$, and integrate with respect to~$z$. If one absorbs the
integration constant of this latter integration into $\tilde{S}_0$,
the result is
\begin{equation}
  \frac{\pa}{\pa w}S(z,w)|_{w=0} = \frac{1}{f(z)^2} - \frac{1}{z^2}
    \quad\Longrightarrow\quad \tilde{S}(Z) = \frac{1}{F(Z)^2}-\frac{1}
    {\tan^2 Z}
\end{equation}
with
\begin{equation}
  F(Z):=f(\tan Z)=f(z)\, .
\end{equation}
Putting everything together, we obtain from~(\ref{eq:Szwsplit}):
\begin{equation}
\begin{split}
  (1+w^2)S(z,w) = & -\int_0^W dZ'\,\tilde{S}(Z') + \frac{1}{2}\int_0^{Z+W}
    dZ'\,\tilde{S}(Z') - \frac{1}{2}\int_0^{Z-W}dZ'\,\tilde{S}(Z') \\
  & {}+\frac{1}{2}\Big(\frac{1}{F(Z+W)}-\frac{1}{\tan(Z+W)}\Big)
    +\frac{1}{2}\Big(\frac{1}{F(Z-W)}-\frac{1}{\tan(Z-W)}\Big) \, .
\end{split}
\end{equation}
This expression should be compared with eq.~(\ref{eq:Szw}). Finally,
a somewhat messy calculation leads to the following result:
\begin{equation}
  \frac{1}{2}\frac{1-F'(Z+W)}{F(Z+W)^2} + \frac{1}{2}\frac{1+F'(Z-W)}
    {F(Z-W)^2} = \frac{F''(W)}{F(W)} + \frac{1-F'(W)^2}{F(W)^2} +
    \frac{F''(W)}{F(Z)-F(W)} + \frac{F'(W)^2}{(F(Z)-F(W))^2}\, .
  \label{eq:surstcond}
\end{equation}
This condition restricts the allowed maps~$f(z)$. A short computation
shows that it is indeed satisfied for the generalized butterfly states
with $F(Z)=\frac{1}{a}\sin aZ$ as well as the wedge states with $F(Z)=
\sfrac{1}{a}\tan aZ$. By a differentiation w.\,r.\,t.\ $Z$ and an
integration over~$W$, eq.~(\ref{eq:surstcond}) can be transformed into
\begin{equation}
  -\frac{F'(Z)F'(W)}{(F(Z)-F(W))^2} = \frac{1-F'(Z+W)}{2F(Z+W)^2} -
    \frac{1+F'(Z-W)}{2F(Z-W)^2}\, . \label{eq:surstcond2}
\end{equation}
The integration constant is fixed by the initial conditions $F(0)=0$,
$F'(0)=1$, and $F''(0)=0$. In this formulation, the left-hand side of
eq.~(\ref{eq:surstcond}) is a bosonic two-point function. It is easy
to see that condition~(\ref{eq:surstcond2}) is a refined version of
the surface state condition~(3.40) in~\cite{Fuchs:2002zz}, cf.\ also
eq.~(3.30) in this reference. From the mathematical point of view, this
condition is a differential equation with deviating arguments.

\section{Solution of the consistency condition} \label{sec:solvcc}
\noindent
In this section, we will solve the surface state condition~%
(\ref{eq:surstcond2}). It will turn out that the general solution is
given by the Jacobi sine function. However, not all maps in the
resulting two-parameter family of solutions define admissible
surface states.

\noindent
{\bf Equivalent ODE.} The general technique will be to derive an
ordinary differential equation from~(\ref{eq:surstcond2}) sharing
all solutions of the consistency equation. It is possible to obtain
two different differential equations with (deviating) arguments $Z$
and $2Z$ from eq.~(\ref{eq:surstcond2}); from the combination of both,
we obtain the desired ordinary differential equation.

On the one hand, by taking into account that $F(Z)$ is an odd function,
the substitution $W\mapsto -W$ in~(\ref{eq:surstcond2}) yields
\begin{equation}
  -\frac{F'(Z)F'(W)}{(F(Z)+F(W))^2} = \frac{-1-F'(Z+W)}{2F(Z+W)^2} +
    \frac{1-F'(Z-W)}{2F(Z-W)^2}\, . \label{eq:surstcond3}
\end{equation}
The sum of eqs.~(\ref{eq:surstcond2}) and~(\ref{eq:surstcond3}) can
be integrated with respect to $Z$ and $W$ to give, respectively:
\begin{align}
  \frac{F'(W)}{F(Z)+F(W)} + \frac{F'(W)}{F(Z)-F(W)} & = \frac{1}{F(Z+W)}
    + \frac{1}{F(Z-W)}\, , \label{eq:DEwdahlp1} \\
  \frac{F'(Z)}{F(Z)+F(W)} - \frac{F'(Z)}{F(Z)-F(W)} & = \frac{1}{F(Z+W)}
    - \frac{1}{F(Z-W)}\, . \label{eq:DEwdahlp2}
\end{align}
Now, the sum of eqs.~(\ref{eq:DEwdahlp1}) and~(\ref{eq:DEwdahlp2}) can be
transformed into
\begin{equation}
  F(Z+W) = \frac{F(Z)^2-F(W)^2}{F(Z)F'(W)-F'(Z)F(W)}\, .
\end{equation}
The limit $W\to Z$ in this equation leads to a further differential
equation with deviating arguments:
\begin{equation}
  F(2Z) = -2\frac{F(Z)F'(Z)}{F(Z)F''(Z) - F'(Z)^2}\, . \label{eq:DEwdaZ2Z_1}
\end{equation}
This is the first differential equation with arguments $Z$ and $2Z$.

On the other hand, the limit $W\to -Z$ in eq.~(\ref{eq:surstcond2})
yields the second equation of this kind,
\begin{equation}
  \frac{1}{2}\frac{F'(Z)^2}{F(Z)^2} = \frac{1+F'(2Z)}{F(2Z)^2} +
    \frac{1}{2}\, F'''(0)\, . \label{eq:DEwdaZ2Z_2}
\end{equation}

Inserting eq.~(\ref{eq:DEwdaZ2Z_1}) into eq.~(\ref{eq:DEwdaZ2Z_2}), we
obtain the desired ordinary differential equation:
\begin{equation}
  \frac{3}{2}\frac{F''(Z)}{F(Z)} - \frac{1}{2}\frac{F'''(Z)}
    {F'(Z)} = F'''(0) =: \frac{c}{2}\, . \label{eq:eqvODE}
\end{equation}
This is consistent for $Z\to 0$. One can already read off at this point
that given a solution~$F(Z)$ to~(\ref{eq:eqvODE}), $\tilde{F}_a(Z):=
\frac{1}{a}F(aZ)$ is also a solution for arbitrary $a\in\C$ (with
$\tilde{F}_0(Z)=Z$ since $F'(0)=1$) obeying the same initial conditions.
We understand the right-hand side as an integration constant~$\frac{c}{2}$.

It is obvious that solutions to~(\ref{eq:surstcond2}) are automatically
solutions to eq.~(\ref{eq:eqvODE}). Indeed, the converse is also true;
as a physicists' proof it may suffice to state that from both equations,
the Taylor coefficients of $F(Z)=Z+\sum_{k=1}^\infty a_k Z^{2k-1}$ have
to satisfy the relations
\begin{align}
  a_3 & = -\sfrac{3}{7}\,a_1^3 + \sfrac{11}{7}\,a_1 a_2\, ,\notag \\
  a_4 & = -\sfrac{2}{7}\,a_1^4 + \sfrac{5}{7}\,a_1^2 a_2 + \sfrac{5}{6}\,
            a_2^2\, , \label{eq:Fouriercond} \\
  a_5 & = -\sfrac{46}{77}\,a_1^3 a_2 + \sfrac{307}{154}\,a_1 a_2^2\, ,
           \notag \\
      & \qquad\vdots \notag 
\end{align}
In order to obtain these conditions from~(\ref{eq:surstcond2}), it is
convenient to expand
\begin{equation}
  G(Z,\b Z) = \frac{F'(Z)F'(\b Z)}{(F(Z)-F(\b Z))^2} +
    \frac{1-F'\big((1+\b)Z\big)} {2F\big((1+\b)Z\big)^2} -
    \frac{1+F'\big((1-\b)Z\big)} {2F\big((1-\b)Z\big)^2}
\end{equation}
(with $W=\b Z$; such a $\b$ exists as long as $Z\neq 0$, the statement
for $Z=0$ follows from continuity) w.\,r.\,t.\ $Z$ and $\b$. Note that this
equivalence is not really necessary for our purposes since we will solve eq.~(\ref{eq:eqvODE}); physical requirements will select two one-parameter
families of functions out of the two-parameter family of solutions to eq.~(\ref{eq:eqvODE}) for which it is easy to check that also the original
consistency condition~(\ref{eq:surstcond2}) is satisfied.

\noindent
{\bf Solving the equivalent ODE.} The ordinary differential equation
can be integrated easily using standard methods: Since eq.~(\ref{eq:eqvODE})
does not contain $Z$ explicitly, we can introduce $y=\frac{dF}{dZ}$, $x=F$
so that $\frac{d}{dZ}=y\frac{d}{dx}$, and~(\ref{eq:eqvODE}) reduces to
\begin{equation}
  \frac{3y y'}{x} - (y'^2 + y y'') = c\, .
\end{equation}
In this paragraph, the primes denote derivatives w.\,r.\,t.\ $x$.
In terms of $g:=y^2$, the latter equation can be rewritten as
\begin{equation}
  \frac{3}{2x} g' - \frac{1}{2} g'' = c\, ,
\end{equation}
with the obvious solution $g(x)=b\,x^4 + \frac{c}{2}\,x^2 + d$, where $b$
and $d$ are integration constants. Substituting back, we obtain
\begin{equation}
  \frac{dF}{dZ}(F) = \pm\sqrt{bF^4 + \sfrac{c}{2}F^2 + d}\, .
  \label{eq:eqvODEsolvhlp}
\end{equation}
The initial condition $\frac{dF}{dZ}(F=0)=1$ implies the choice of the
plus sign and $d=1$. If we introduce
\begin{equation}
  \Delta_+ := -\frac{c}{4b} + \sqrt{\frac{c^2}{16b^2}-\frac{1}{b}}
    \qquad\text{and}\qquad
  \Delta_- := -\frac{c}{4b} - \sqrt{\frac{c^2}{16b^2}-\frac{1}{b}}\, ,
\end{equation}
this equation for $d$ translates into $b\Delta_+\Delta_-=1$. Then,
eq.~(\ref{eq:eqvODEsolvhlp}) can be integrated,
\begin{equation}
  Z=\int_0^{F(Z)/\sqrt{\Delta_+}}\frac{\sqrt{\Delta_+}\,dt}{\sqrt{(1-t^2)
    (1-\sfrac{\Delta_+}{\Delta_-}t^2)}}\, ,
\end{equation}
and this is with $\sqrt{b\Delta_-}=:a$ and $\frac{\Delta_+}{\Delta_-}=:
k^2$ the defining equation~(\ref{eq:SNdef}) of the Jacobi sine function
with modulus $k$. Thus,
\begin{equation}
  F(Z) = \frac{1}{a}\sn(aZ,k)
\end{equation}
is the promised solution to eq.~(\ref{eq:eqvODE}). One can easily
check that its Fourier coefficients satisfy eqs.~(\ref{eq:Fouriercond}).
This proves that all twist-even $\k$-diagonal surface states are of the
form
\begin{equation}
  f(z) = \frac{1}{a}\sn(a\tan^{-1} z, k)
  \label{eq:sol}
\end{equation}
with some (so far unrestricted) complex constants $a, k$. One should note
that this is already a fair amount of reduction of degrees of freedom:
We started out from a twist-even $\k$-diagonal squeezed state with two
independent functions $S_{ee}(\k)$ and $S_{oo}(\k)$ and ended up with a
two-parameter family of solutions to our consistency conditions. We will
see, however, that physical constraints even select two one-parameter
solutions of the form~(\ref{eq:sol}).

\section{$\k$-Diagonal surface states} \label{sec:kdiagss}
\noindent
In this section, we will analyze which maps of the form~(\ref{eq:sol})
actually determine surface states. Namely in general, the maps $f(z)$
defining a surface state are subject to several constraints which are not
automatically satisfied by all solutions to our consistency condition.
This analysis will require some knowledge of Jacobian elliptic
functions; their basic properties are summarized in appendix~\ref{sec:Jac}.

\noindent
{\bf Conditions on $f$.} There are several conditions a map $f(z)$
defining a surface state has to satisfy. E.\,g.~\cite{Schnabl:2002ff},
\begin{enumerate}
  \item[(i)] $f$ should be meromorphic,
  \item[(ii)] it should map the real axis to the real axis,
  \item[(iii)] it should map the unit upper half disk (in the $z$-plane)
        into some region of the upper half plane and
  \item[(iv)] it should be one-to-one (at least) for $|z|<1$.
\end{enumerate}
The last condition guarantees that $f(z)$ furnishes a good local
coordinate around the puncture at $f(0)$ on the surface~$\S$.
Although the computation of correlation functions requires $f(z)$
only to be defined locally around the origin, the possibility to
glue the surface necessitates a map from the full upper half disk
into~$\S$~\cite{Rastelli:2001uv}. These requirements will
suffice to reduce the number of possible surface states in the
twist-even $\k$-diagonal sector tremendously.

As a doubly periodic function, $f(z)$ in eq.~(\ref{eq:sol}) is
meromorphic; thus, condition~(i) entails no further conditions
on the candidate functions in eq.~(\ref{eq:sol}). Since
\begin{equation}
  \frac{1}{a}\sn(aZ,k)\stackrel{!}{=}\ov{\frac{1}{a}\sn(aZ,k)}
  = \frac{1}{\ab}\sn(\ab\Zb,\kb) = \frac{1}{\ab}\sn(\ab Z,\kb)
  \label{eq:solreal}
\end{equation}
for $Z$ real,\footnote{The bar denotes complex conjugation; and $Z$
abbreviates $\tan^{-1} z$ (as above). Note that the latter function
maps the real line to the real line (more explicitly, to the interval
$(-\frac{\pi}{4},\frac{\pi}{4})$) and the upper half plane to the
region $\text{Re }Z\leq\frac{\pi}{2}$, $\text{Im }Z\geq 0$.}
requirement~(ii) has the obvious solutions $a=\pm\ab$, $k=\pm\kb$
(where the signs can be chosen independently), i.\,e.~$a$ and~$k$
are restricted to be real or purely imaginary. However, because
of~(\ref{eq:SNktmo}), the right-hand side equals
\begin{equation}
  \frac{1}{\ab}\sn(\ab Z,\kb) = \frac{1}{\ab\kb}\sn(\ab\kb Z,\kb^{-1})\, .
  \label{eq:solreal2}
\end{equation}
Since eq.~(\ref{eq:sol}) features no periodicity in $a$ or $k$
one can read off that the only other solution to condition~(ii) are
functions with arbitrary $a\neq 0$, $k=\pm\frac{\ab}{a}\in S^1$.%
\footnote{Here, $S^1$ denotes the unit sphere in the complex $k$-plane.
Note that due to the symmetry $\sn(u,k)=\sn(u,-k)$, we need not
distinguish between the latter two cases. Similarly, a distinction
between $a$ and $-a$ in eq.~(\ref{eq:sol}) is irrelevant since~$\sn$
is odd in its first argument.\label{fn:ama}} We will now scrutinize
properties~(iii) and~(iv) for these cases.

\begin{figure}[t]
\parbox{7.8cm}{
\resizebox{7.2cm}{7.2cm}{\includegraphics{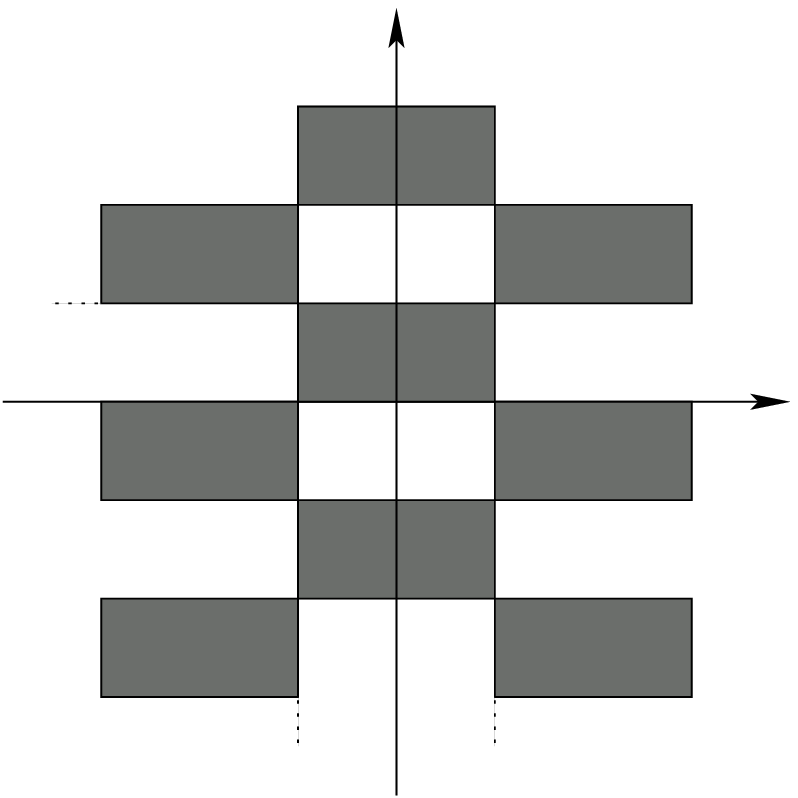}}
\setlength{\unitlength}{1cm}
\begin{picture}(0,0)
  \put(-3.0,0.2){\makebox(0,0)[c]{$K$}}
  \put(-4.8,0.2){\makebox(0,0)[c]{$-K$}}
  \put(-7.3,4.5){\makebox(0,0)[c]{$K'$}}
  \put(-0.1,3.3){\makebox(0,0)[c]{$U$}}
  \put(-4.2,7.0){\makebox(0,0)[c]{$V$}}
  \put(-7.0,7.0){\makebox(0,0)[c]{(a)}}
\end{picture}}
\hfill
\parbox{7.8cm}{\vspace{5mm}
\resizebox{7.6cm}{5.6cm}{\includegraphics{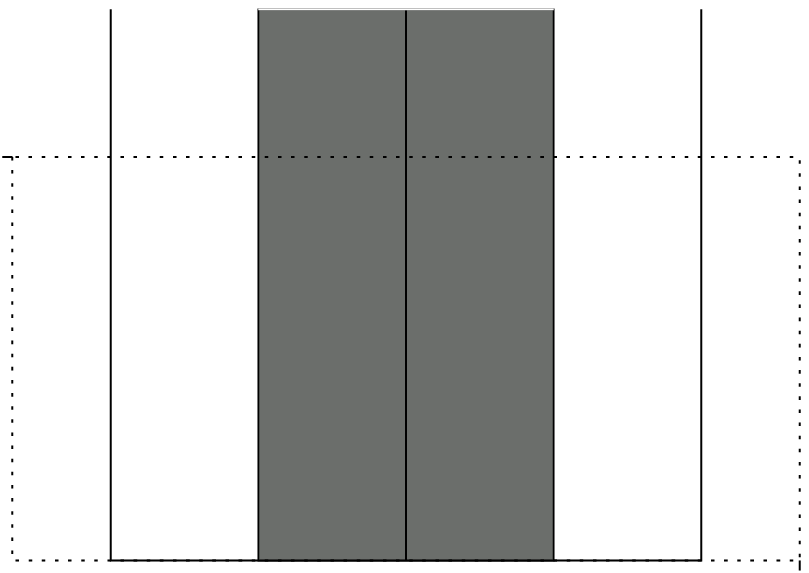}}
\setlength{\unitlength}{1cm}
\begin{picture}(0,0)
  \put(-6.95,-0.2){\makebox(0,0)[c]{$-a\pi/2$}}
  \put(-5.55,-0.2){\makebox(0,0)[c]{$-a\pi/4$}}
  \put(-2.55,-0.2){\makebox(0,0)[c]{$a\pi/4$}}
  \put(-1.15,-0.2){\makebox(0,0)[c]{$a\pi/2$}}
  \put(-8.05, 4.1){\makebox(0,0)[c]{$K'$}}
  \put(-0.25, -0.2){\makebox(0,0)[c]{$K$}}
  \put(-7.3,6.3){\makebox(0,0)[c]{(b)}}
\end{picture}}
\caption{\label{fig:arkr}The situation where $k$ and $a$ are real:
In (a), the shaded region denotes the part of the $UV$-plane
which is mapped to the upper half plane. The tile including the
origin from (a) is enlarged (and plotted with dashed lines) in (b).
Here, the shaded region denotes the image of the local coordinate
patch under the map $z\mapsto a\tan^{-1} z$; condition~(iii) demands
that it fits into the dashed tile.}
\end{figure}
\noindent
{\bf Case~1: $k$, $a$ real.} Since $\tan^{-1}$ maps the unit upper half
disk into the region~$|\text{Re }Z|\leq\frac{\pi}{4}$, $\text{Im }Z\geq 0$,
property~(iii) requires that $\frac{1}{a}\sn(aZ,k)$ maps this region
into the upper half plane again. Using eqs.~(\ref{eq:SNadd}),
(\ref{eq:SNiu}), (\ref{eq:CNiu}) and~(\ref{eq:DNiu}) as well as
$U:=\text{Re }aZ$, $V:=\text{Im }aZ$, we derive that
\begin{equation}
  \frac{1}{a}\sn(U+\ic V) = \frac{1}{a}\frac{\sn(U,k)\dn(V,k')+\ic\sn(V,k')
    \cn(V,k')\cn(U,k)\dn(U,k)}{\cn^2(V,k')+k^2\sn^2(V,k')\sn^2(U,k)}\, .
  \label{eq:SNdecomp}  
\end{equation}
Here, $k'=\sqrt{1-k^2}$ denotes the complementary modulus.
By inspection of~(\ref{eq:SNdecomp}), we find that the function on
the left-hand side has positive imaginary part for \mbox{$\sn(V,k')\cn(V,k')
\cn(U,k)\dn(U,k)\geq 0$}, i.\,e., $U\in [(2m-1)K,(2m+1)K]$ and $V\in [(m+2n)K',(m+2n+1)K']$. The situation is illustrated in figure~%
\ref{fig:arkr}(a). Obviously, property~(iii) here leads to the
requirement that
\begin{equation}
  K\geq\frac{a\pi}{4}\, ,\qquad K'=\infty\, . \label{eq:sincond}
\end{equation}
The second condition in~(\ref{eq:sincond}) fixes $k=0$, then, the first
requires $a\in [0,2]$.\footnote{As explained in the footnote on page~%
\pageref{fn:ama}, we may restrict the analysis to positive~$a$.} For
$k=0$, the Jacobi sine reduces to the ordinary sine function; therefore,
this is the butterfly family with
\begin{equation}
  f(z) = \frac{1}{a}\sin(a\tan^{-1} z)\qquad \text{with } a\in [0,2]\, .
  \label{eq:butfam}
\end{equation}
In particular, it contains the sliver state with $f(z)=\tan^{-1} z$ for
$a=0$, the butterfly state with $f(z)=\frac{z}{\sqrt{1+z^2}}$ for $a=1$,
and the nothing state with $f(z)=\frac{z}{1+z^2}$ for $a=2$.
Condition~(iv) is automatically satisfied.

\begin{figure}[t]
\parbox{7.8cm}{
\resizebox{7.2cm}{7.2cm}{\includegraphics{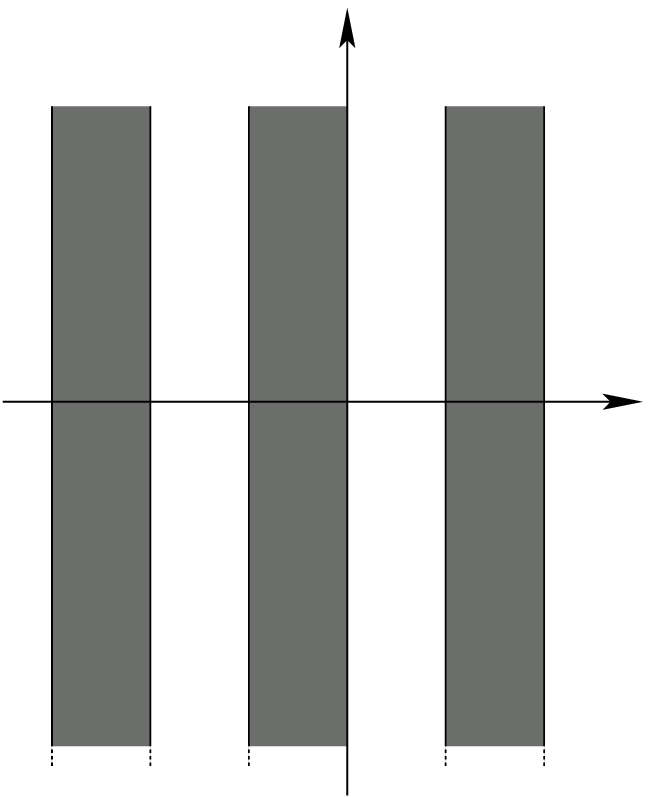}}
\setlength{\unitlength}{1cm}
\begin{picture}(0,0)
  \put(-0.25,3.3){\makebox(0,0)[c]{$U$}}
  \put(-3.95,7.1){\makebox(0,0)[c]{$V$}}
  \put(-1.35,0.05){\makebox(0,0)[c]{$4K$}}
  \put(-2.45,0.05){\makebox(0,0)[c]{$2K$}}
  \put(-4.75,0.05){\makebox(0,0)[c]{$-2K$}}
  \put(-5.85,0.05){\makebox(0,0)[c]{$-4K$}}
  \put(-6.95,0.05){\makebox(0,0)[c]{$-6K$}}
  \put(-6.7,7.1){\makebox(0,0)[c]{(a)}}  
\end{picture}}
\hfill
\parbox{7cm}{
\resizebox{5.5cm}{7.6cm}{\includegraphics{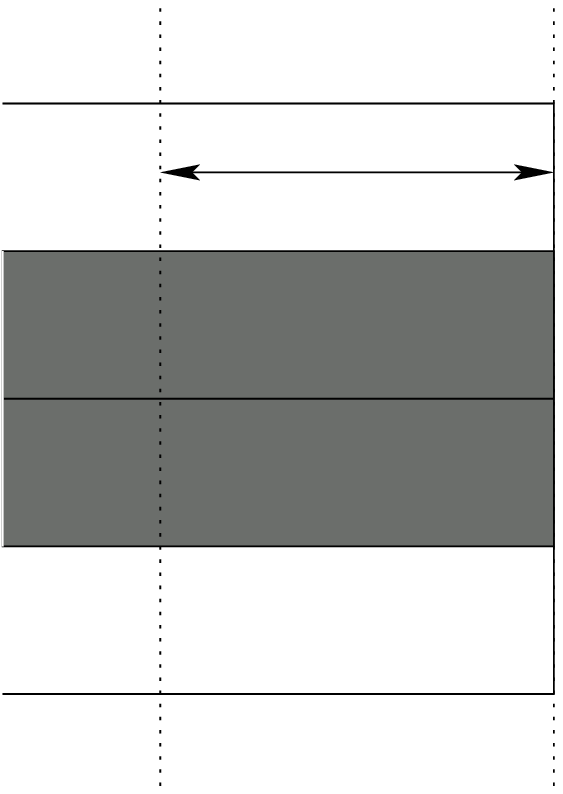}}
\setlength{\unitlength}{1cm}
\begin{picture}(0,0)
  \put(0.4, 0.95){\makebox(0,0)[c]{$-a\pi/2$}}
  \put(0.4, 2.35){\makebox(0,0)[c]{$-a\pi/4$}}
  \put(0.3, 5.2){\makebox(0,0)[c]{$a\pi/4$}}
  \put(0.3, 6.6){\makebox(0,0)[c]{$a\pi/2$}}
  \put(-2.2, 6.15){\makebox(0,0)[c]{$2K$}}
  \put(-5.7,7.2){\makebox(0,0)[c]{(b)}}
\end{picture}}
\caption{\label{fig:aikr}The situation where $k$ is real and $a$ is
imaginary: In (a), the shaded region denotes the part of
the $UV$-plane that is mapped to the upper half plane. The tile
including the origin from (a) is enlarged (and plotted with dashed
lines) in (b). Here, the shaded region denotes the image of the
local coordinate patch under the map $z\mapsto a\tan^{-1} z$.}
\end{figure}
\noindent
{\bf Case 2: $k$ real, $a$~imaginary.} Again, we start with condition~%
(iii). For the analysis of this case, we can recycle
eq.~(\ref{eq:SNdecomp}). It is easy to see that $F(Z)$ maps the region
where $\sn(U,k)\dn(V,k')\leq 0$ to the upper half plane. This region is
depicted in figure~\ref{fig:aikr}(a). Denoting $a=\ic a_0$ with $a_0$
real, we see that $Z=\frac{1}{a_0}(V-\ic U)$, i.\,e.\ the positive
imaginary half-axis of~$Z$ extends along the negative $U$-axis in figure~\ref{fig:aikr}. We read off that
\begin{equation}
  K = \infty\qquad\Rightarrow\qquad k=1\, .
\end{equation}
For $k=1$, the Jacobi sine reduces to the $\tanh$-function, so that
the solution may be rewritten as
\begin{equation}
  f(z) = \frac{1}{a_0}\tan(a_0 \tan^{-1} z)\qquad\text{with }a_0\in
    [0,2]\, . \label{eq:wdgfam}
\end{equation}
For $a_0=\frac{2}{n}$ with $n\in\Na$, this defines the customary wedge
states, including the sliver state for $a_0=0$ and the identity state
for $a_0=2$ with $f(z)=\frac{z}{1-z^2}$. We have bounded $a_0$ from
above in order to ensure condition~(iv); for $a_0\nearrow 2$, the image
of the unit disk already fills the upper half plane; for $a_0>2$, the
map is no langer one-to-one.

\begin{figure}[h]
\parbox{7.2cm}{
\resizebox{7.1cm}{7.1cm}{\includegraphics{arkrpip.eps}}
\setlength{\unitlength}{1cm}
\begin{picture}(0,0)
  \put(7.15,3.65){\makebox(0,0)[c]{$U$}}
  \put(3.25,7.5){\makebox(0,0)[c]{$V$}}
  \put(4.4, 0.6){\makebox(0,0)[c]{$\widetilde{K}$}}
  \put(2.5, 0.6){\makebox(0,0)[c]{$-\widetilde{K}$}}
  \put(0.05, 4.95){\makebox(0,0)[c]{$2\widetilde{K}'$}}
  \put(0.2,7.5){\makebox(0,0)[c]{(a)}}
\end{picture}}
\hfill
\parbox{7.2cm}{
\resizebox{7.1cm}{7.1cm}{\includegraphics{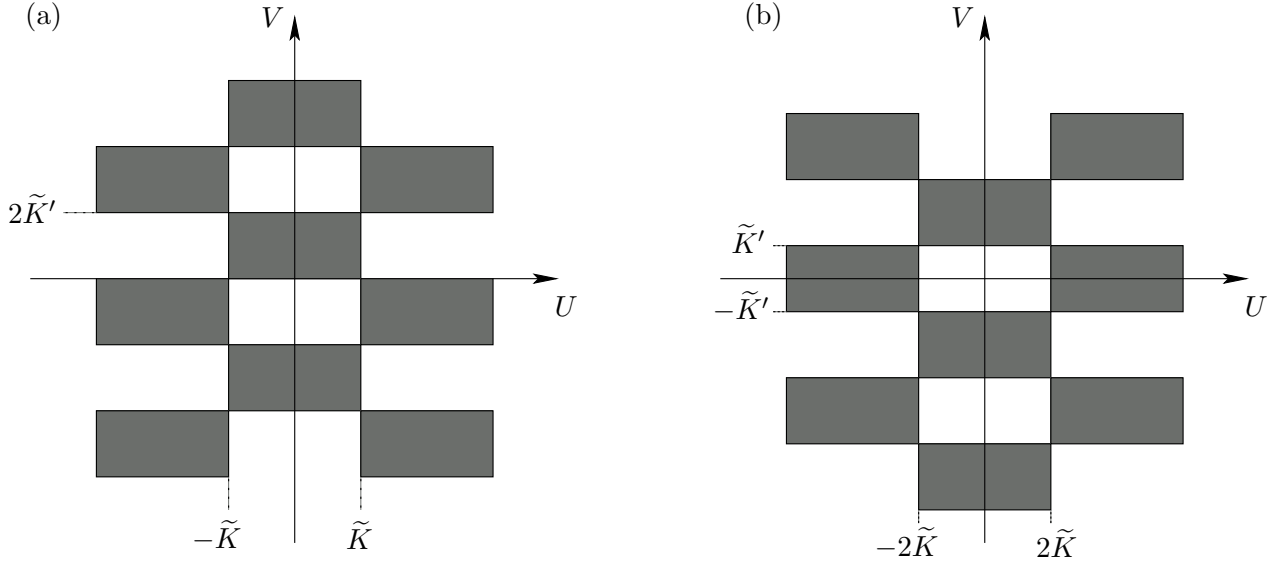}}
\setlength{\unitlength}{1cm}
\begin{picture}(0,0)
  \put(7.15,3.65){\makebox(0,0)[c]{$U$}}
  \put(3.25,7.5){\makebox(0,0)[c]{$V$}}
  \put(4.5, 0.5){\makebox(0,0)[c]{$2\widetilde{K}$}}
  \put(2.5, 0.5){\makebox(0,0)[c]{$-2\widetilde{K}$}}
  \put(0.3, 3.65){\makebox(0,0)[c]{$-\widetilde{K}'$}}
  \put(0.4, 4.55){\makebox(0,0)[c]{$\widetilde{K}'$}}
  \put(0.6,7.5){\makebox(0,0)[c]{(b)}}
\end{picture}}
\vspace{-2mm}
\caption{\label{fig:ki}The situation where $k$ is imaginary: In (a),
$a$ is real; the shaded region denotes the part of the $UV$-plane
that is mapped to the upper half plane. In (b), $a$ is imaginary;
the meaning of the shaded region is as in (a).}
\end{figure}

\noindent
{\bf Case 3: $k$ imaginary, $a$ real.} We first scrutinize condition~%
(iii). An imaginary value of the elliptic modulus, $k=\ic k_0$, can be
mapped to a real value $\kt:=k_0/\sqrt{1+k_0^2}\in [0,1)$.
Introducing the abbreviations $\at:=\sqrt{1+k_0^2}\,a$, $\kt':=\sqrt{1-\kt^2}=1/\sqrt{1+k_0^2}$, $\Ut:=\text{Re }\at Z$,
and $\Vt:=\text{Im }\at Z$, we can use eqs.~(\ref{eq:SNik}),
(\ref{eq:SNadd}), and~(\ref{eq:DNadd}) to find the following decomposition
of~$f(z)$ into real and imaginary parts:
\begin{equation}
\label{eq:kidecomp}
\begin{split}
  F(Z) = & \,\frac{1}{\at}\,g(\Ut,\Vt,\kt)\,\big(\sn(\Ut,\kt)
    \dn(\Vt,\kt') + \ic\cn(\Ut,\kt)\dn(\Ut,\kt)\sn(\Vt,\kt')
    \cn(\Vt,\kt')\big) \\
  & \times\big(\dn(\Ut,\kt)\cn(\Vt,\kt')\dn(\Vt,\kt')+ 
    \ic\kt^2\sn(\Ut,\kt)\cn(\Ut,\kt)\sn(\Vt,\kt')\big)
\end{split}
\end{equation}
with some real and positive definite function~$g(\Ut,\Vt,\kt)$. For
real~$a$, the imaginary part is positive exactly where $\cn(\Ut,\kt)
\sn(\Vt,\kt')\dn(\Vt,\kt')$ is positive. This region is depicted
in figure~\ref{fig:ki}(a). Just as above, one can read off that
\begin{equation}
  \widetilde{K}:=K(\kt)\geq\frac{a\pi}{4}\, ,\qquad
  \widetilde{K}':=K(\kt')=\infty\, .
\end{equation}
The latter condition fixes~$k=0$ which was already dealt with in
case~1. There are no new solutions in this situation.

\noindent
{\bf Case 4: $k$, $a$ imaginary.} With the same notation as above,
we may read off from eq.~(\ref{eq:kidecomp}) that the imaginary
part of $F(Z)$ is positive exactly where
\begin{equation}
  \sn(\Ut,\kt)\dn(\Ut,\kt)\cn(\Vt,\kt')\big( -\dn^2(\Vt,\kt')+
  \kt^2\cn^2(\Ut,\kt)\sn^2(\Vt,\kt')\big) \geq 0\, .
\end{equation}
Due to eqs.~(\ref{eq:SN2CN2})--(\ref{eq:CN2DN2}), the expression in
brackets equals $-\kt^2\sn^2(\Ut,\kt)-\dn^2(\Ut,\kt)\cn^2(\Vt,\kt')$.
Thus, $F(Z)$ maps the region where $\sn(\Ut,\kt)\dn(\Ut,\kt)
\cn(\Vt,\kt')\leq 0$ to the upper half plane. This situation is
illustrated in figure~\ref{fig:ki}(b); obviously, there are no values
of~$\kt$ such that $f(z)$ maps the unit upper half disk completely
into the upper half plane again.

\noindent
{\bf Case 5: $a\in\C\backslash\{0\}$, $k=\pm\ab/a$.} The last case
is the situation where~$k$ takes values in the unit circle. The
analysis in this case is a slightly more involved than before. We
restrict the discussion to the case $k=\frac{\ab}{a}$; the other
case with $k=-\frac{\ab}{a}$ can be treated analogously. We
already saw in eqs.~(\ref{eq:solreal}) and~(\ref{eq:solreal2}) that
with~$k=\pm\ab/a$, 
\begin{equation}
  \frac{1}{a}\sn\big(aX,\sfrac{\ab}{a}\big)\in\R \label{eq:SNXre}
\end{equation}
for $X\in\R$; similarly, it is trivial to see that
\begin{equation}
  \frac{1}{a}\sn\big(\ic aY,\sfrac{\ab}{a}\big)\in\ic\,\R
  \label{eq:SNiYim}
\end{equation}
for $Y\in\R$. Since furthermore, for $k=1$, $\cn(u,k)=\ov{\dn(u,k)}$,
eq.~(\ref{eq:SNadd}) yields the following imaginary part:
\begin{equation}
  \text{Im }\frac{1}{a}\sn\big(a(X+\ic Y)\big) =
  \frac{-\sfrac{\ic}{a}\sn(\ic aY)|\cn(aX)|^2}{1-\sfrac{|a|^4}{a^4}
    \sn^2(aX)\sn^2(\ic aY)} \, , 
\end{equation}
where the elliptic modulus is always~$\sfrac{\ab}{a}$. Because of~%
(\ref{eq:SNXre}) and~(\ref{eq:SNiYim}), the denominator is positive;
thus, the imaginary part is positive exactly where $-\sfrac{\ic}{a}
\sn(\ic aY)$ is. A careful analysis using the zeroes and singularities
given in appendix~\ref{sec:Jac} and the conjugation properties~%
(\ref{eq:Kpconj}) and~(\ref{eq:Kconj}) shows that the singularities
of the Jacobi sine function bound the real values of~$Y$ for which the
imaginary part of~$-\sfrac{\ic}{a}\sn(\ic aY)$ is positive. The
intervals for which this is so are (for $k=\text{e}^{\ic \a}$):
\begin{equation}
  -\frac{\ic}{a}\sn(\ic aY)\geq 0\qquad\text{for }
  \begin{cases}
    Y\in [0,\frac{1}{a}K') & \text{if }-\frac{\pi}{2}<\a<
      \frac{\pi}{2}\, ,\\
    Y\in [0,\frac{2}{\ic a}K+\frac{1}{a}K') & \text{if }\frac{\pi}
      {2}<\a<\pi\, ,\\
    Y\in [0,\frac{2}{\ic a}K-\frac{1}{a}K') & \text{if }\pi<\a<
      \frac{3\pi}{2}\, .
  \end{cases}
\end{equation}
As above, the elliptic modulus is always~$\sfrac{\ab}{a}$.
Since condition~(iii) requires the upper bound of these
intervals to be infinite this case only leads (asymptotically)
to the solution $k=\text{e}^{\ic\pi}=-1$ (i.\,e., the sliver, which
was already discussed in case~1).

This concludes the proof that there are no twist-even $\k$-diagonal
surface states apart from the butterfly family~(\ref{eq:butfam})
and a trivial extension of the wedge states admitting arbitrary
angles, cf.\ eq.~(\ref{eq:wdgfam}).

\section{Conclusions}\label{sec:concl}
\noindent
In this paper, we have derived and analyzed a consistency condition on
twist-even $\k$-diagonal surface states. The general solution was found
to be a two-parameter family of (candidate) maps defining a surface state,
$\frac{1}{a}\sn(a\tan^{-1} z,k)$. This family contains the well-known
butterfly family as well as an extension of the family of wedge states
to arbitrary angles; and the conditions that the meromorphic function
$f(z)$ map the real line to itself and the unit upper half disk one-to-one
to some region of the upper half plane were enough to show that these are
in fact the only admissible maps to define a surface state.

In particular, this result demonstrates that surface state projectors
which do not satisfy the sufficient condition that their boundary touches
the midpoint of the open string (i.\,e., $f(\ic)=\infty$ in upper half
plane coordinates) are particularly rare -- at least in the subsector of
twist-even $\k$-diagonal surface state projectors, we only found the
identity state. All other projectors share this singular property.

Furthermore, this result completes the classification of $\k$-diagonal
projectors (in a fermionic first order system) commenced in~\cite{Ihl:2003fw}. Besides, it also proves the conjecture made in~\cite{Fuchs:2002zz} that
the projectors $P_{2m}$ defined by $f(z)=z(1-(-z^2)^m)^{-1/2m}$ for $m>1$
are not contained in the ${\cal H}_{\k^2}$ subalgebra. Hitherto, this
conjecture had only been checked numerically.

It is conceivable that the strategy pursued here can be extended to
the general case without the restriction to the twist-even subsector.
The hope that underlies this approach is that a better knowledge of
the algebraic structures of the star algebra will help to pave the way
to solving the equation of motion of string field theory.

\subsubsection*{Acknowledgements}
\noindent
I would like to thank Alexander Kling, Yuji Okawa, Martin Schnabl,
Wati Taylor, David Tong, and Barton Zwiebach for discussions. Furthermore,
I would like to thank Matthias Ihl for encouragement as well as a
collaboration at an intermediate stage. This work was supported
by a fellowship within the postdoc program of the German Academic
Exchange Service (DAAD).

\begin{appendix}
\section{Jacobian elliptic functions}\label{sec:Jac}
\noindent
In this appendix, we present the definitions and properties of the
Jacobian elliptic functions used throughout the text. A good reference
for some of the formulas is~\cite{GR1}.

\noindent
{\bf Definitions.} The Jacobian elliptic functions (of which we need
here only $\sn$, $\cn$, and $\dn$) are doubly-periodic functions defined
as the (analytic continuation of the) inverse of elliptic integrals,
\begin{align}
  u & = \int_0^{\sn u} \frac{dt}{\sqrt{(1-t^2)(1-k^2 t^2)}}\, ,
    \label{eq:SNdef}\\
  u & = \int_1^{\cn u} \frac{dt}{\sqrt{(1-t^2)(k'^2-k^2 t^2)}}\, , \\
  u & = \int_1^{\dn u} \frac{dt}{\sqrt{(1-t^2)(t^2-k'^2)}}\, .
\end{align}
Here, $k$ denotes the modulus, $k'=\sqrt{1-k^2}$ is the complementary
modulus. Note that a number of incompatible conventions for the modulus
are in common use, e.\,g., $k$ is often replaced by its square.
If the modulus is obvious, it is mostly omitted (as above); otherwise,
we denote it explicitly by $\sn(u,k)$ etc. For $k=0$, $\sn$ and $\cn$
reduce to the ordinary trigonometric functions, $\sn(u,0)=\sin u$ and
$\cn(u,0)=\cos u$ (and $\dn(u,0)=1$). For $k=1$, $\sn(u,1)=\tanh u$ and
$\cn(u,1)=\dn(u,1)=\text{sech }u$.

\noindent
{\bf Complete elliptic integrals.} The periodicity and zeroes of the
Jacobian elliptic functions can be expressed in terms of the complete
elliptic integral of the first kind,
\begin{equation}
  K := K(k) = \int_0^1 \frac{dx}{\sqrt{(1-x^2)(1-k^2 x^2)}}\, ,
\end{equation}
where the modulus~$k$ is often omitted. This is the value~$u$ where the
Jacobi sine function equals~1. It is customary to set
\begin{equation}
  K' := K(k')\, .
\end{equation}
In general, we have
\begin{equation}
  \ov{K(k)} = K(\kb)\, ,\qquad \ov{K'(k)} = K'(\kb)\, ,
\end{equation}
where the bar denotes complex conjugation. This ensures that the period
parallelogram of the Jacobian elliptic functions specified in the next
paragraph is a rectangle. The elliptic integral of the first kind has a
branch cut from~1 to~$\infty$.

Particular values are $K(0)=\frac{\pi}{2}$ and $K(1)=\infty$.
For the analysis in section~\ref{sec:kdiagss}, we derive some further
properties for $|k|=1$: Setting $k=\text{e}^{\ic\a}=\frac{\ab}{a}$ and
being careful with the branchcuts, we can check that
\begin{equation}
  K'\big(\sfrac{1}{k}\big) =
  \begin{cases}
    k\,K'(k) &\text{for }-\frac{\pi}{2}<\a<\frac{\pi}{2}\, , \\
    -k\,K'(k) &\text{for }\frac{\pi}{2}<\a<\frac{3\pi}{2}\, , \\
  \end{cases}
\end{equation}
and
\begin{equation}
  K\big(\sfrac{1}{k}\big) =
  \begin{cases}
    k(K(k)-\ic K'(k)) &\text{for }0<\a<\frac{\pi}{2}\, ,\\
    -k(K(k)+\ic K'(k)) &\text{for }\frac{\pi}{2}<\a<\pi\, ,\\
    -k(K(k)-\ic K'(k)) &\text{for }\pi<\a<\frac{3\pi}{2}\, ,\\
    k(K(k)+\ic K'(k)) &\text{for }\frac{3\pi}{2}<\a< 2\pi\, .
  \end{cases}
\end{equation}
Therefore, we obtain the conjugation properties
\begin{equation}
  \ov{\frac{1}{a}K'(k)} = \frac{1}{\ab}K'\big(\sfrac{1}{k}\big)
  \begin{cases}
    \frac{1}{a}K'(k) &\text{for }-\frac{\pi}{2}<\a<\frac{\pi}{2}\, , \\
    -\frac{1}{a}K'(k) &\text{for }\frac{\pi}{2}<\a<\frac{3\pi}{2}
  \end{cases}
  \label{eq:Kpconj}
\end{equation}
and
\begin{equation}
  \ov{\frac{1}{\ic a}K(k)} = -\frac{1}{\ic a}K\big(\sfrac{1}{k}\big) =
  \begin{cases}
    -\frac{1}{\ic a}K(k)+\frac{1}{a}K'(k) &\text{for }
      0<\a<\frac{\pi}{2}\, ,\\
    \frac{1}{\ic a}K(k)+\frac{1}{a}K'(k) &\text{for }
      \frac{\pi}{2}<\a<\pi\, ,\\
    \frac{1}{\ic a}K(k)-\frac{1}{a}K'(k) &\text{for }
      \pi<\a<\frac{3\pi}{2}\, ,\\
    -\frac{1}{\ic a}K(k)-\frac{1}{a}K'(k) &\text{for }
      \frac{3\pi}{2}<\a< 2\pi\, .  
  \end{cases}
  \label{eq:Kconj}
\end{equation}

\noindent
{\bf Symmetries, periodicities, zeroes, and poles.} The Jacobi sine
function is an odd function of its first argument, the Jacobi cosine
function as well as the delta function are even w.\,r.\,t.\ to~$u$.
All three functions are even in~$k$. For $|u|<K'$, the power series
expansion takes the form:
\begin{align}
  \sn u & = u - \frac{1+k^2}{3!}u^3 + \frac{1+14k^2+k^4}{5!}u^5 -
    \frac{1+135k^2+135k^4+k^6}{7!}u^7 + \ldots\, , \\
  \cn u & = 1 - \frac{1}{2!}u^2 + \frac{1+4k^2}{4!}u^4 -
    \frac{1+44k^2+16k^4}{6!} u^6 + \ldots\, ,\\
  \dn u & = 1-\frac{k^2}{2!}u^2 + \frac{k^2(4+k^2)}{4!}u^4 -
    \frac{k^2(16+44k^2+k^4)}{6!}u^6 + \ldots\, .
\end{align}
A closed form for the Taylor coefficients is unknown. The Jacobian
elliptic functions are not periodic in~$k$; the periods, zeroes, and
poles in~$u$ are located at:% \\[2mm]
\begin{center}
\begin{tabular}{|l|c|c|c|}\hline
  & Periods & Zeroes & Poles\\ \hline
  $\sn u$ & $4mK+2nK'\ic$ & $2mK+2nK'\ic$ & $2mK+(2n+1)K'\ic$\\
  $\cn u$ & $4mK+2n(K+K'\ic)$ & $(2m+1)K+2nK'\ic$ & $2mK+(2n+1)K'\ic$\\
  $\dn u$ & $2mK+4nK'\ic$ & $(2m+1)K+(2n+1)K'\ic$ & $2mK+(2n+1)K'\ic$\\
  \hline
\end{tabular}% \\[3mm]
\end{center}
Here, it is understood that $m, n\in\Z$. One can read off from
\begin{align}
  \sn(u+2mK+2nK'\ic,k) & = (-1)^m\sn(u,k)\, , \\
  \cn(u+2mK+2nK'\ic,k) & = (-1)^{m+n}\cn(u,k)\, , \\
  \dn(u+2mK+2nK'\ic,k) & = (-1)^n\dn(u,k)\, ,    
\end{align}
and $\sn(0,k)=0$, $\cn(K,k)=0$ that the Jacobi sine function changes
sign only along the real axis whereas the Jacobi cosine function does
so along both axes.

In general, we have
\begin{equation}
  \ov{\sn(u,k)} = \sn(\ub,\kb)\, ,\quad \ov{\cn(u,k)} = \cn(\ub,\kb)\, ,
  \quad \ov{\dn(u,k)} = \dn(\ub,\kb)\, .
\end{equation}
For $|k|=1$, $\cn$ and $\dn$ are complex conjugate to each other in
the sense that $\cn(u,k)=\ov{\dn(u,k)}$.

\noindent
{\bf Useful formulas.} In order to check which parts of the complex
plane are mapped to the upper half plane, we need the following useful
identities:
\begin{align}
  \sn(\ic u, k) & = \ic\frac{\sn(u,k')}{\cn(u,k')}\, , \label{eq:SNiu}\\
  \cn(\ic u, k) & = \frac{1}{\cn(u,k')}\, , \label{eq:CNiu}\\
  \dn(\ic u, k) & = \frac{\dn(u,k')}{\cn(u,k')} \label{eq:DNiu}\\
  \intertext{as well as}
  \sn(u,k) & = k^{-1}\sn(ku,k^{-1})\, , \label{eq:SNktmo}\\
  \cn(u,k) & = \dn(ku,k^{-1})\, , \\
  \dn(u,k) & = \cn(ku,k^{-1})
  \intertext{and}
  \sn(u,\ic k) & = \frac{1}{\sqrt{1+k^2}}\frac{\sn(\sqrt{1+k^2}u,
    k/\sqrt{1+k^2})}{\dn(\sqrt{1+k^2}u, k/\sqrt{1+k^2})}\, ,
    \label{eq:SNik}\\
  \cn(u,\ic k) & = \frac{\sn(\sqrt{1+k^2}u, k/\sqrt{1+k^2})}
    {\dn(\sqrt{1+k^2}u, k/\sqrt{1+k^2})}\, , \\
  \dn(u,\ic k) & = \frac{1}{\dn(\sqrt{1+k^2}u, k/\sqrt{1+k^2})}\, .\end{align}

\noindent
{\bf Addition formulas.} In analogy to the addition formulas for
the trigonometric functions, we have:
\begin{align}
  \sn(u\pm v) & = \frac{\sn u\cn v\dn v\pm\sn v\cn u\dn u}
    {1-k^2\sn^2 u\sn^2 v} \, , \label{eq:SNadd}\\
  \cn(u\pm v) & = \frac{\cn u\cn v\mp\sn u\sn v\dn u\dn u}
    {1-k^2\sn^2 u\sn^2 v} \, , \\
  \dn(u\pm v) & = \frac{\dn u\dn v\mp k^2\sn u\sn v\cn u\cn v}
    {1-k^2\sn^2 u\sn^2 v} \, . \label{eq:DNadd}
\end{align}

\noindent
{\bf Relation among the elliptic functions.} The Jacobian elliptic
functions are related by:
\begin{align}
  \sn^2 u + \cn^2 u & = 1\, , \label{eq:SN2CN2}\\
  k^2\sn^2 u + \dn^2 u & = 1\, , \\
  k^2\cn^2 u + k'^2 & = \dn^2 u\, , \label{eq:CN2DN2}\\
  \cn^2 u + k'^2\sn^2 u & = \dn^2 u\, .
\end{align}
\end{appendix}

\end{document}